\documentclass[prb,twocolumn,amssymb,showpacs,amsmath,nobibnotes,superscriptaddress,floatfix,aps]{revtex4}
\usepackage{graphicx}
\usepackage{bm}

\begin{document}
\title {Vortex states in iron-based superconductors with collinear antiferromagnetic cores}

\author{Hong-Min Jiang}
\affiliation{National Laboratory of Solid State of Microstructure
and Department of Physics, Nanjing University, Nanjing 210093,
China} \affiliation{Department of Physics and Center of Theoretical
and Computational Physics, The University of Hong Kong, Pokfulam
Road, Hong Kong, China}
\author{Jian-Xin Li}
\affiliation{National Laboratory of Solid State of Microstructure
and Department of Physics, Nanjing University, Nanjing 210093,
China}
\author{Z. D. Wang} \affiliation{Department of Physics and Center of Theoretical and
Computational Physics, The University of Hong Kong, Pokfulam Road,
Hong Kong, China}

\date{\today}

\begin{abstract}
Magnetism in the FeAs stoichiometric compounds and its interplay
with superconductivity in vortex states are studied by
self-consistently solving the BdG equations based on a two-orbital
model with including  the on-site interactions between electrons in
the two orbitals. It is revealed that for the parent compound,
magnetism is caused by the strong Hund's coupling, and the Fermi
surface topology aids to select the spin-density-wave (SDW) pattern.
The superconducting (SC) order parameter with
$s_{\pm}=\Delta_{0}\cos(k_{x})\cos(k_{y})$ symmetry is found to be
the most favorable pairing for both the electron- and hole-doped
cases, while the local density-of-states (LDOS) exhibits the
characteristic of nodal gap for the former and full gap for the
latter. In the vortex state, the emergence of the field-induced SDW
depends on the strength of the Hund's coupling and the Coulomb
repulsions. The field-induced SDW gaps the finite energy contours on
the electron and hole pocket sides, leading to the dual structures
with one reflecting the SC pairing and the other being related to
the SDW order. These features can be discernable in STM measurements
for identifying the interplay between the field-induced SDW order
and the SC order around the core region.
\end{abstract}

\pacs{74.20.Mn, 74.25.Ha, 74.25.Jb, 74.72.Bk}
 \maketitle

\section{introduction}
The recently discovered iron arsenide
superconductors,~\cite{kami1,xhchen1,zaren1,gfchen1,wang1} which
display superconducting transition temperature as high as more than
50K, appear to share a number of general features with high-$T_{c}$
cuprates, including the layered structure and proximity to a
magnetically ordered state.~\cite{kami1,cruz1,jdong} The accumulated
evidences have subsequently established a fact that the parent
compounds are generally poor metal and undergo structure and
antiferromagnetic (AFM) spin-density-wave (SDW) transitions below
certain temperatures.~\cite{cruz1,klau1} Elastic neutron scattering
experiments have shown the antiferromagnetic order is collinear and
has a wavevector $(\pi,0)$ or $(0,\pi)$ in the unfolded Brillouin
zone corresponding to a unit cell with only one Fe atom per unit
cell.~\cite{cruz1} Either chemical doping or/and pressure suppresses
the AFM SDW instability and eventually results in the emergence of
superconductivity.~\cite{kami1,taka1} The novel magnetism and
superconducting properties in these compounds have been a great spur
to recent
researches.~\cite{zhao1,chen1,goto1,luet1,qsi1,ccao1,djsingh,yao1,huwz,daikou,qhwang,ragh1}

The relation between magnetism and superconductivity and the origin
of magnetic order have attracted significant attentions in the
current research on FeAs superconductors. Discrepancies exist in the
experimental results, i.e., whether the superconductivity and
antiferromagnetic order are well separated or they can coexist in
the underdoped region of the phase diagram, and how they coexist if
they happen to do so. For example, there is no overlap between those
two phases in CeFeAsO$_{1-x}$F$_{x}$~\cite{zhao1}, while the
coexistence of the two phases was observed in a narrow doping range
in SmFeAsO$_{1-x}$F$_{x}$~\cite{liudrew}, and in a broader range in
Ba$_{1-x}$K$_{x}$Fe$_{2}$As$_{2}$~\cite{chen1,goto1}. Even for the
same LaFeAsO$_{1-x}$F$_{x}$ system, different experiments display
conflicting results. It was reported that before the orthorhombic
SDW phase is completely suppressed by doping, superconductivity has
already appeared at low temperatures~\cite{kami1}, while it was also
observed experimentally that superconductivity appears after the SDW
is completely suppressed~\cite{luet1}. As for the origin of the SDW
phase, two distinct types of theories have been proposed: local
moment antiferromagnetic ground state for strong
coupling,~\cite{qsi1} and itinerant ground state for weak
coupling.~\cite{ccao1,djsingh,yao1,ragh1} The detection of the local
moment seems to question the weak coupling scenario, but the
metallic-like (or bad metal) nature as opposed to a correlated
insulator as in cuprates renders the strong coupling theories
doubtable.~\cite{huwz} More recently,  a compromised scheme was
adopted: the SDW instability is assumed to result from the coupling
of itinerant electrons with local moment, namely, neither the Fermi
surface nesting nor the local moment scenario alone is able to
account for it.~\cite{daikou}

Although many research efforts have been already made to identify
the existence of magnetic order and its origin as well as the
relationship with superconductivity, there have been fewer studies
on vortex states of the systems. While the interplay between
magnetism and superconductivity has been yet to be experimentally
clarified, the superconducting critical temperature $T_{c}$ reaches
its maximum value after the antiferromagnetic spin order is
completely suppressed in the materials, indicating the competition
nature between AFM SDW instability and superconductivity. At this
stage, it is valuable and interesting to investigate vortex states
in the family of FeAs compounds, mainly considering that the
magnetic order may arise naturally when the superconducting order is
destroyed by the magnetic vortex. Therefore, one can perform local
tunneling spectroscopic probes in vortex states to understand
profoundly the interplay between magnetic order and
superconductivity.

In this paper, we investigate magnetism in the FeAs stoichiometric
compounds, and the interplay between it and superconductivity upon
doping in vortex states by self-consistently solving the BdG
equations based on the two-orbital model with including the on-site
interactions between electrons in the two orbitals. It is shown that
for the parent compound, magnetism is caused by the strong Hund's
coupling, and the Fermi surface topology aids to select the SDW
ordering pattern. The SDW results in the pseudogap-like feature at
the Fermi level in the LDOS. It is found that the SC order parameter
with $s_{\pm}=\Delta_{0}\cos(k_{x})\cos(k_{y})$ symmetry is the most
favorable pairing at both the electron- and hole-doped sides, while
the LDOS exhibits the characteristic of nodal gap for the former and
full gap for the latter. In the vortex states, the emergence of the
field-induced SDW order depends heavily on the strength of the
Hund's coupling and the Coulomb repulsions.  The coexistence of the
field-induced SDW order and SC order around the core region is
realized due to the fact that the two orders emerge at different
energies. The corresponding LDOS at the core region displays a kind
of dual structures, with one reflecting the SC pairing and the other
being related to the SDW order.

The paper is organized as follows. In Sec. II, we introduce the
model Hamiltonian and carry out analytical calculations. In Sec.
III, we present numerical calculations and discuss the results. In
Sec. IV, we make remarks and conclusion.

\section{THEORY AND METHOD}
We start with an effective two-orbital model~\cite{ragh1} that takes
only the iron $d_{xz}$ and $d_{yz}$ orbitals into account. By
assuming an effective attraction that causes the superconducting
pairing and including the possible interactions between the two
orbitals' electrons, one can construct an effective model to study
the vortex physics of the iron-based superconductors in the mixed
state:
\begin{eqnarray}
H=H_{0}+H_{pair}+H_{int}.
\end{eqnarray}
The first term is a tight-binding model
\begin{eqnarray}
H_{0}=&&-\sum_{ij,\alpha\beta,\sigma}e^{i\varphi_{ij}}t_{ij,\alpha\beta}
c^{\dag}_{i,\alpha,\sigma}c_{j,\beta,\sigma}
  \nonumber \\
   &&-\mu\sum_{i,\alpha,\sigma}c^{\dag}_{i,\alpha,\sigma}c_{i,\alpha,\sigma},
\end{eqnarray}
which describes the electron effective hoppings between sites $i$
and $j$ of the Fe ions on the square lattice, including the intra-
($t_{ij,\alpha\alpha}$) and inter-orbital ($t_{ij,\alpha,\beta},
\alpha\neq\beta$) hoppings with the subscripts $\alpha$, $\beta$
($\alpha,(\beta)=1,2$ for $xz$ and $yz$ orbital, respectively)
denoting the orbitals and $\sigma$ the spin.
$c^{\dag}_{i,\alpha\sigma}$ creates an $\alpha$ orbital electron
with spin $\sigma$ at the site $i$ ($i\equiv(i_{x},i_{y})$), and
$\mu$ is the chemical potential. The magnetic field is introduced
through the Peierls phase factor $e^{i\varphi_{ij}}$ with
$\varphi_{ij}=\frac{\pi}{\Phi_{0}}\int^{r_{i}}_{r_{j}}\mathbf{A(r)}\cdot
d\mathbf{r}$, where $A=(-Hy, 0, 0)$ stands for the vector potential
in the Landau gauge and $\Phi_{0}=hc/2e$ is the superconducting flux
quantum. The hopping integrals are chosen as to capture the essence
of the density function theory (DFT) results.~\cite{gxu1} Taking the
hopping integral between the $d_{yz}$ orbitals $|t_{1}|=1$ as the
energy unit, we have,
\begin{eqnarray}
t_{i,i\pm \hat{x},xz,xz}=&&t_{i,i\pm \hat{y},yz,yz}=t_{1}=-1.0 \nonumber\\
 t_{i,i\pm\hat{y},xz,xz}=&&t_{i,i\pm \hat{x},yz,yz}=t_{2}=1.3 \nonumber\\
 t_{i,i\pm\hat{x}\pm\hat{y},xz,xz}=&&t_{i,i\pm\hat{x}\pm
 \hat{y},yz,yz}=t_{3}=-0.9 \nonumber\\
t_{i,i+\hat{x}-\hat{y},xz,yz}=&&t_{i,i+\hat{x}-\hat{y},yz,xz}
 =t_{i,i-\hat{x}+\hat{y},xz,yz} \nonumber\\
=&&t_{i,i+\hat{x}-\hat{y},yz,xz}=t_{4}=-0.85 \nonumber\\
t_{i,i+\hat{x}+\hat{y},xz,yz}=&&t_{i,i+\hat{x}+\hat{y},yz,xz}
 =t_{i,i-\hat{x}-\hat{y},xz,yz} \nonumber\\
=&&t_{i,i-\hat{x}-\hat{y},yz,xz}=-t_{4}.
\end{eqnarray}
Here, $\hat{x}$ and $\hat{y}$ denote the unit vector along the $x$
and $y$ direction, respectively.

The second term accounts for the superconducting pairing.
Considering that a main purpose here is to address the interplay
between the SC and magnetism in the vortex state for the FeAs-based
superconductors, we take a phenomenological form for the pairing
interaction,
\begin{eqnarray}
H_{pair}=&&\sum_{i\neq
j,\alpha\beta}V_{ij}(\Delta_{ij,\alpha\beta}c^{\dag}_{i,\alpha\uparrow}c^{\dag}_{j,\beta\downarrow}
+h.c.)
\end{eqnarray}
with $V_{ij}$ as the strengths of effective attractions.

The third term represents the interactions between
electrons,~\cite{oles1}
\begin{eqnarray}
H_{int}=&& U\sum_{i,\alpha}n_{i,\alpha\uparrow}n_{i,\alpha\downarrow}\nonumber\\
&&
+U^{'}\sum_{i,\alpha<\beta,\sigma}n_{i,\alpha,\sigma}n_{i,\beta,\bar{\sigma}}
\nonumber\\
&&+(U^{'}-J)\sum_{i,\alpha<\beta,\sigma}
n_{i,\alpha,\sigma}n_{i,\beta,\sigma}
\nonumber\\
&&+J^{'}\sum_{i,\alpha<\beta}(c^{\dag}_{i,\alpha\uparrow}c^{\dag}_{i,\alpha\downarrow}c_{i,\beta\downarrow}
c_{i,\beta\uparrow}\nonumber\\
&&+c^{\dag}_{i,\alpha\uparrow}c^{\dag}_{i,\beta\downarrow}c_{i,\alpha\downarrow}
c_{i,\beta\uparrow}+h.c.),
\end{eqnarray}
which includes the intra-orbital (inter-orbital) Coulomb repulsion
$U$ ($U^{'}$), the Hund's rule coupling $J$ as well as the
inter-orbital Cooper pairing hopping term $J^{'}$.

After the Hartree-Fock decomposition of the on-site interaction
term, one arrives at the Bogoliubov-de Gennes equations in the mean
field approximation for this model Hamiltonian
\begin{eqnarray}
\sum_{j,\alpha<\beta}\left(
\begin{array}{cccc}
H_{ij,\alpha\alpha,\sigma} &
\tilde{\Delta}_{ij,\alpha\alpha} & H_{ij,\alpha\beta,\sigma} & \Delta^{\ast}_{ii,\beta\alpha} \\
\tilde{\Delta}^{\ast}_{ij,\alpha\alpha} &
-H^{\ast}_{ij,\alpha\alpha,\bar{\sigma}} &
\Delta_{ii,\alpha\beta} & -H^{\ast}_{ij,\alpha\beta,\bar{\sigma}} \\
H_{ij,\alpha\beta,\sigma} & \Delta^{\ast}_{ii,\alpha\beta} &
H_{ij,\beta\beta,\sigma} & \tilde{\Delta}_{ij,\beta\beta} \\
\Delta_{ii,\beta\alpha} & -H^{\ast}_{ij,\alpha\beta,\bar{\sigma}} &
\tilde{\Delta}^{\ast}_{ij,\beta\beta} &
-H^{\ast}_{ij,\beta\beta,\bar{\sigma}}
\end{array}
\right)\nonumber\\
\times\left(
\begin{array}{cccc}
u^{n}_{j,\alpha,\sigma} \\
v^{n}_{j,\alpha,\bar{\sigma}} \\
u^{n}_{j,\beta,\sigma} \\
v^{n}_{j,\beta,\bar{\sigma}}
\end{array}
\right)= E_{n}\left(
\begin{array}{cccc}
u^{n}_{i,\alpha,\sigma} \\
v^{n}_{i,\alpha,\bar{\sigma}} \\
u^{n}_{i,\beta,\sigma} \\
v^{n}_{i,\beta,\bar{\sigma}}
\end{array}
\right),
\end{eqnarray}
where,
\begin{eqnarray}
H_{ij,\alpha\alpha,\sigma}=&&-e^{i\varphi_{ij}}t_{ij,\alpha\alpha}+\delta_{ij}
[U\langle n_{i,\alpha,\bar{\sigma}}\rangle+ \nonumber\\
&&U^{'}\langle
n_{i,\beta(\beta\neq\alpha),\bar{\sigma}}\rangle+(U^{'}-J)\langle n_{i,\beta(\beta\neq\alpha),\sigma}\rangle
\nonumber\\ &&-\mu] \nonumber\\
H_{ij,\alpha\beta(\beta\neq\alpha),\sigma}=&&-e^{i\varphi_{ij}}t_{ij,\alpha\beta(\beta\neq\alpha)}
\nonumber\\
\tilde{\Delta}_{ij,\alpha\alpha}=&&\Delta_{ij,\alpha\alpha}+
\Delta^{\ast}_{ii,\beta\beta(\beta\neq\alpha)}.
\end{eqnarray}
$u^{n}_{j,\alpha,\sigma}$ ($u^{n}_{j,\beta,\bar{\sigma}}$),
$v^{n}_{j,\alpha,\sigma}$ ($v^{n}_{j,\beta,\bar{\sigma}}$) are the
Bogoliubov quasiparticle amplitudes on the $j$-th site with
corresponding eigenvalues $E_{n}$.

The pairing amplitude and electron densities are obtained through
the following self-consistent equations,~\cite{chen-wang}
\begin{eqnarray}
\Delta_{ij(i\neq
j),\alpha\alpha}=&&\frac{V}{4}\sum_{n}(u^{n}_{i,\alpha,\sigma}
v^{n\ast}_{j,\alpha,\bar{\sigma}}
+v^{n\ast}_{i,\alpha,\bar{\sigma}}u^{n}_{j,\alpha,\sigma})\times \nonumber\\
&&\tanh(\frac{E_{n}}{2k_{B}T})
\nonumber\\
\Delta_{ii,\alpha\beta}=&&\frac{J}{4}\sum_{n}(u^{n}_{i,\alpha,\sigma}
v^{n\ast}_{i,\beta,\bar{\sigma}}
+v^{n\ast}_{i,\beta,\bar{\sigma}}u^{n}_{i,\alpha,\sigma})\times \nonumber\\
&&\tanh(\frac{E_{n}}{2k_{B}T})
\nonumber\\
n_{i,\alpha,\uparrow}=&&\sum_{n}|u^{n}_{i,\alpha,\uparrow}|^{2}f(E_{n}) \nonumber\\
n_{i,\alpha,\downarrow}=&&\sum_{n}|v^{n}_{i,\alpha,\downarrow}|^{2}[1-f(E_{n})].
\end{eqnarray}

The electronic structure associated with the SDW and the vortex
states, namely, the local density-of-states (LDOS),
$N(\mathbf{r}_{i},E)$ is calculated by
\begin{eqnarray}
N(\mathbf{r}_{i},E)=&&
-\sum_{n,\alpha}[|u_{i,\alpha,\uparrow}^{n}|^{2}f^{'}(E_{n}-E) \nonumber\\
&&+|v_{i,\alpha,\downarrow}^{n}|^{2} f^{'}(E_{n}+E)].
\end{eqnarray}
where, $f^{'}(E)$ is the derivative of the Fermi-Dirac distribution
function with respect to energy.

In numerical calculations, the undoped case is determined by the
equality of the area enclosed by the electron- and hole-pocket in
the unfolded Brillouin zone, which leads to
$n_{h}=\sum_{i,\alpha,\sigma}(n_{\alpha,\sigma})/(N_{x}N_{y})=2$.
The Coulomb interactions $U, U^{'}, J, J^{'}$ are expected to
satisfy the conventional relation $U^{'}=U-2J$ and
$J^{'}=J$.~\cite{yao1,wern1}  In the literatures, $U\sim 0.2W-0.5W$
and $J\sim 0.09W$ are expected.~\cite{ccao1,yao1} Here, $W$ is the
energy bandwidth, which is $12.4|t_{1}|$ in our case. This gives
rise to $J\sim |t_{1}|$~\cite{haule} and $U\sim 2.2J-5.5J$. We have
found numerically that the results presented here are not subject to
the qualitative changes in the intermediate coupling range $U\approx
3J\sim 4J$, where the ground state is an AFM
metal.~\cite{lorenz,rongyu} In the following, the typical result
with $U=3.5J$ will be presented. We take $V_{ij}=0$ for the normal
state. In the SC state, $V_{ij}$ is chosen to give a short coherence
length of a few lattice spacing being consistent with
experiments.~\cite{takeshita} We use $V_{ij}=V=2.0$, $\mu=1.75$
($\mu=1.13$), which gives rise to the filling factor
$n=\sum_{i,\alpha,\sigma}(n_{\alpha,\sigma})/(N_{x}N_{y})=2.2$
($n=1.8$) and the coherent peak of the SC order parameter in the DOS
being at $\Delta_{max}\simeq0.4$. Thus, we estimate the coherence
length $\xi_{0}\sim E_{F}a/|\Delta_{max}|~\cite{ydzhu}\sim4a$. Due
to this short coherence length, presumably the system will be a
type-II superconductor. The unit cell with size $N_{x}\times
N_{y}=40\times20$ and the number of such unit cells $M_{x}\times
M_{y}=10\times20$ are used in the numerical calculations. In view of
these parameters, we estimate the upper critical field
$B_{c2}\sim130T$. Therefore, the model calculation is particularly
suitable for the iron-based type-II superconductors such as
CaFe$_{1-x}$Co$_{x}$AsF, Eu$_{0.7}$Na$_{0.3}$Fe$_{2}$As$_{2}$ and
FeTe$_{1-x}$S$_{x}$ , where the typical coherence length $\xi_{0}$
deduced from the experiments is of a few lattice
spacing~\cite{takeshita} and the upper critical field achieves as
high as dozens of Tesla.

\section{results and discussion}

\subsection{SDW phase in the absence of the magnetic field}
In the absence of a magnetic field and pairing term, we obtain the
collinear AFM SDW at the half filling. Fig. 1(a) is the typical
result with $J=0.96$ for the real space distribution of the moment
$M_{i}$ defined as $M_{i}=\sum_{\alpha}M_{i,\alpha}$ with
$M_{i,\alpha}=\frac{1}{2}(\langle
n_{i,\alpha,\uparrow}\rangle-\langle
n_{i,\alpha,\downarrow}\rangle)$ being the spin order defined on the
$\alpha$ orbital. As can be seen in Fig. 1(a), the real space
distribution of $M_{i}$ antiferromagnetically alines along the $x$
direction but ferromagnetically along the $y$ direction. In Fig.
1(b) Fourier transformation of $M_{i}$ gives an SDW order with wave
vector $Q=(\pi,0)$, which is consistent with experimental results in
the undoped systems.~\cite{cruz1,chen2} [For another initial input
parameters, the degenerate configuration of $M_{i}$ with wave vector
$(0,\pi)$ can be obtained.] We note that the emergence of magnetic
order is heavily dependent on the Hund's coupling strength $J$. For
$J=0$, the magnetic ordered phase does not exist even with very
large $U$ and $U^{'}$. Therefore, magnetism itself is generated by
the strong Hund's coupling, whereas the Fermi surface topology aids
to select the ordering pattern.~\cite{mazi2} This is a reminiscent
of the spin freezing phase found in a three-orbital model relevant
to transition metal oxide SrRuO$_{3}$,~\cite{wern1} and may be a
common feature respecting the magnetic order origin in multiple
orbital systems involving the Hund's coupling interaction.

In Fig. 1(c), we plot the LDOS $N(\mathbf{r}_{i},E)$ in the SDW
state at a site with positive $M_{i}$, i.e., the spin-up site
labeled $A$ in Fig. 1(a). The electronic structure in the SDW state
displays a clear pseudogap-like feature with a heavily depressed but
nonvanishing density-of-state (DOS) at the Fermi energy, pointing to
the metallic magnetic ordered state. The magnetic order derived
pseudogap-like feature is consistent with the experimental
observation of partial gaps in the SDW state of the parent
compounds~\cite{wzhu1} and may account for the pseudogap feature in
several experiments.~\cite{liuzhao,boysai}

The pseudogap feature comes from a fact that when the SDW order with
the wave vector $Q$ is involved, there will be gaps on those parts
of the Fermi surface which are best connected by the wave vector
$Q$, while those who are not connected by the wave vector $Q$ remain
untouched, leading to the partial gaps in the SDW states of the
parent compounds.~\cite{hsie1} We make this point more clearly in
Fig. 1(d), in which the spectral weight distribution
$I(k)=\int_{\varepsilon_{F}-\Delta w}^{\varepsilon_{F}+\Delta
w}A(k,\omega)d\omega$ is shown. Here, $A(k,\omega)$ is the
single-particle spectral function and $\Delta w$ is an integration
window. As shown, both the electron and hole Fermi pockets are
partially gaped.

\subsection{Configuration of the order parameters}
In the search for the most favorable pairing symmetry, we consider
all possible singlet pairings, including the extended $s$- and
$d$-wave symmetries, between the nearest, next-nearest, and
third-nearest neighbor (NN, NNN, TNN) sites, as shown in Figs.
2(a)-2(c). The pairing amplitude of the $s$-wave symmetry has the
same sign along the $x$ and $y$ directions for the NN or TNN sites
pairing, resulting in the $k$ dependent pairing form
$\Delta(k)=\Delta_{0}[\cos(k_{x})+\cos(k_{y})]$ for the NN sites
pairing and $\Delta(k)=\Delta_{0}[\cos(2k_{x})+\cos(2k_{y})]$ for
the TNN sites pairing, respectively; and the same sign along the
$x=y$ and $x=-y$ directions for the NNN sites pairing, resulting in
the $k$ dependent pairing form
$\Delta(k)=\Delta_{0}[\cos(k_{x})\cos(k_{y})]$. The $d$-wave
pairing, on the other hand, has amplitude $+\Delta_{0}$ along the
$x$ direction and $-\Delta_{0}$ along the $y$ direction for the NN
or TNN sites pairing, resulting in the $k$ dependent pairing form
$\Delta(k)=\Delta_{0}[\cos(k_{x})-\cos(k_{y})]$ for the NN sites
pairing and $\Delta(k)=\Delta_{0}[\cos(2k_{x})-\cos(2k_{y})]$ for
the TNN sites pairing, respectively; and $+\Delta_{0}$ along $x=y$
direction and $-\Delta_{0}$ along $x=-y$ direction for the NNN sites
pairing, resulting in the $k$ dependent pairing form
$\Delta(k)=\Delta_{0}[\sin(k_{x})\sin(k_{y})]$.

The introduction of pairing interaction suppresses the SDW order
completely on both the electron- and hole-doped sides, and leads to
the homogeneous SC order in real space. We carry out extensive
calculations, and find that in the reasonable doping range the most
favorable pairing symmetry is the intra-orbital pairing between NNN
sites,
\begin{eqnarray}
\Delta_{i,i+\hat{x}+\hat{y},\alpha\alpha}=&&\Delta_{i,i-\hat{x}-\hat{y},\alpha\alpha}
=\nonumber\\
\Delta_{i,i+\hat{x}-\hat{y},\alpha\alpha}=&&\Delta_{i,i-\hat{x}+\hat{y},\alpha\alpha}
=\Delta_{i,\alpha\alpha},
\end{eqnarray}
which leads to the $s_{\pm}$-wave pairing
$s_{\pm}=\Delta_{0}\cos(k_{x})\cos(k_{y})$, being consistent with
that obtained before.~\cite{yao1,mazihu,dhfcch} Then, the
superconducting order parameter $\Delta_{i}$ is expressed as,
\begin{eqnarray}
\Delta_{i}=\frac{1}{2}\sum_{\alpha}\Delta_{i,\alpha\alpha}.
\end{eqnarray}
For the choice of $V=2.0$ in this paper, one gets the amplitude
$\Delta_{i}\sim0.12$ for the $s_{\pm}$ SC order .

\subsection{Vortex states}
When a magnetic field is applied, the SC order parameter around the
vortex core is suppressed, so that the system may be driven into a
vortex state. We find that there exists a critical Hund's coupling
value $J_{c}$ separating the regimes of two kinds of vortex states
associated respectively with and without the field-induced SDW
order. In the following, we address these two regimes in detail.

\subsubsection{The vortex state without the field-induced SDW order}
The vortex state without the field induced SDW order is stable when
$J$ is less than $J_{ce}=0.9$ on the electron-doped side with
$n=2.2$ and less than $J_{ch}=1.25$ on the hole-doped side with
$n=1.8$, respectively. Typical results on the nature of the vortex
state are displayed in Fig. 3 for $n=2.2$ with the Hund's coupling
$J=0.85$, for which no magnetic order is induced. As shown in Fig.
3(a), each unit cell accommodates two superconducting vortices each
carrying a flux quantum $hc/2e$. The SC order parameter $\Delta_{i}$
vanishes at the vortex core center and recovers its bulk value at
the core edge with radium $\xi_{1}$ on the scale of coherent length
$\xi_{0}$.

In Figs. 3(b) and 3(c) we plot the LDOS as a function of energy at
the vortex core center in the absence of the field-induced magnetic
order for electron-doped case with $n=2.2$ and hole-doped case with
$n=1.8$, respectively. For comparison, we have also displayed the
LDOS at the midpoint between two nearest neighbor vortices along the
$x$ direction, which resembles that for the bulk system. As seen
from Figs. 3(b) and 3(c), when $J=0.85$ for which no local SDW order
is induced, the LDOS at the core center shows a single resonant peak
within the SC gap edge for both the electron- and hole-doped cases,
which is similar to that reported by other authors for the cuprates
high-$T_{c}$ superconductors in the vortex state.~\cite{yongwang}
However, the differences are obvious with respect to the position of
the resonant peak and the line shape of the bulk LDOS between the
electron- and hole-doped cases in despite of the same SC pairing
symmetry considered here. More specifically, for the electron-doped
case, the position of the in-gap resonant peak is almost at the
Fermi level and the bulk system exhibits the $V$-shaped LDOS curve,
the typical characteristics which indicate a nodal SC gap. However,
for the hole-doped case, the resonant peak deviates from the Fermi
level to a higher energy and the bulk system exhibits the $U$-shaped
LDOS curve, from which the conclusion for a full SC gap can be made.

The notable differences can be qualitatively understood as follows:
The Fermi surface of the FeAs superconductors consists of hole Fermi
surfaces around the $\Gamma$-point at $(k_{x}, k_{y}) = (0, 0)$
forming the hole-pocket, and the electron Fermi surfaces around the
$M_{1,2}$ points at $(\pi, 0)$ and $(0, \pi)$ forming the
electron-pocket, respectively. Both Fermi pockets change their size
upon doping as depicted in Fig. 3(d), where only the relevant
electron-pockets are displayed. The size of the electron-pocket
enlarges and approaches to the nodal line of the $s_{\pm}$ SC gap
with electron doping while shrinks and deviates from the nodal line
with hole doping. Thus the low energy quasiparticles in the SC phase
show the nodal behavior in the electron-doped system and nodeless
behavior in the hole-doped system. This may explain the discrepancy
observed in experiments concerning the pairing symmetry, where the
conclusion for the nodal gap were obtained on the electron-doped
LnFeAsO (Ln stands for the rare earth elements)
samples~\cite{mushr,grluca,maahma} and a dominant full gap feature
was found on the hole-doped (Ba,Sr)$_{1-x}$K$_{x}$Fe$_{2}$As$_{2}$
systems~\cite{limu,wzhu1} in the measurement of the thermal and
transport properties.

\subsubsection{The vortex state with the field-induced SDW order}
As $J$ increases to about $J_{ce}=0.9$ on the electron-doped side
with $n=2.2$ and $J_{ch}=1.25$ on the hole-doped side with $n=1.8$,
the SDW order is induced around the core region. Fig. 4 displays the
vortex structure with $J=0.96$ for the electron-doped case, where
the local magnetic order is induced around the vortex core, as shown
in Fig. 4(b) which presents the spatial distribution of the local
SDW order as defined in Sec. III A. As seen in Fig. 4(a), the vortex
core expands further with a radium $\xi_{2}$ compared with that in
Fig. 3(a). Meanwhile, the maximum strength of $M_{i}$ appears at the
vortex core center and decays with a scale of $\xi_{2}$ to zero into
the superconducting region, depicting a competition nature between
SC and magnetic orders as observed in experiment~\cite{prat1}.

In this case, it is shown that there is no obvious splitting of the
in-gap bound state peak though the peak intensity is suppressed
heavily in the LDOS for both the electron- and hole-doped cases, as
shown in Figs. 4(c) and 4(d). In addition to the in-gap state peak,
an additional peak structure below the Fermi energy appears for the
electron-doped case, while two peaks situate separately around and
above the Fermi energy for the hole-doped case. These features are
dramatically different from the high-$T_{c}$ cuprates in vortex
states with the field-induced antiferromagnetic order, where the
in-gap resonant peak of the core bound state is split into two peaks
sitting symmetrically about the Fermi energy.~\cite{maggio}

Simply, one can analyze the present vortex state in the following
way: there are two factors that play a role in the physics around
the vortex core region. One is the pure SC of vortex state without
SDW in the magnetic field, while the other is the SDW state without
the SC order for the doped case. It is known that doping destroys
the nesting property between parts of the Fermi surface on the
electron- and hole-pocket, due to the size change of the hole- and
electron-pocket upon doping compared with that in the undoped case.
However, as depicted in Figs. 5(c) and 5(d), the SDW wave vector $Q$
now connects the finite energy contour for the doped case, resulting
in the gap-like feature below (above) Fermi energy in the LDOS shown
in Fig. 5(a) (Fig. 5(b)) for electron (hole) doped case. Combination
of the in-gap resonant peak in the vortex state without SDW and the
SDW-induced finite energy gap feature produce a kind of dual
structures of the LDOS for the finite doped case, ie., the in-gap
bound state peak reflecting the SC pairing and the other peak
structure being related to the SDW order.

\vspace*{-.2cm}
\section{remarks and conclusion}
Clarification of the interplay between magnetism and
superconductivity is a key step toward the understanding of the
underlying physics of the Fe-based high-$T_{c}$ superconductors.
Although the competition nature between them has been identified in
both classes of materials, some still show a coexistence of
them.~\cite{liudrew,chen1,goto1,prat1} Competition between the AFM
SDW and SC is natural in FeAs compounds when one considers that both
originate from the multiple Fe $d$ conduction bands, but quest for
the mechanism of the coexistence of them is shown to be more
challenged. Recently, an incommensurate SDW state with wave vector
$Q^{'}=Q\pm q$ has been proposed to account for the coexistence of
the AFM SDW and superconductivity at finite
doping~\cite{cvet1,voro1}. In such a state, the mismatch between the
electron and hole Fermi pocket at finite doping is compensated by
the incommensurate wave vector $q$, leading to the inferior
"nesting" between the electron and hole Fermi pocket and allowing
for the coexistence of magnetism and superconductivity. This
mechanism only works at the doping level near the AFM instabilities,
where the mismatch between the electron and hole Fermi pocket is
small. Here, we show that the field-induced SDW at the doping level
being far from the AFM instabilities is commensurate with the same
wave vector $Q$ as in the undoped case, and it gaps the finite
energy contour on the electron and hole pocket sides. Therefore, at
optimal doping, the field-induced SDW and the SC order around the
core region may associate with the DOS at different energies,
allowing them to coexist.

In conclusion, we have studied magnetism in the FeAs stoichiometric
compounds and the interplay between it and superconductivity upon
doping in the vortex state by self-consistently solving the BdG
equations based on the two-orbital model including the on-site
interactions between electrons in the two orbitals. It has been
shown that for the parent compound, magnetism is caused by the
strong Hund's coupling, and the Fermi surface topology aids to
select the SDW ordering pattern. The SDW results in the
pseudogap-like feature at the Fermi level in the LDOS. The SDW order
is completely suppressed upon the introduction of the SC
interaction. We have also found that the SC order parameter with
$s_{\pm}=\Delta_{0}\cos(k_{x})\cos(k_{y})$ symmetry is the most
favorable pairing at both the electron- and hole-doped sides, while
the LDOS exhibits the characteristic of nodal gap for the former and
full gap for the latter. In vortex states, the emergence of the
field-induced SDW order depends heavily on the strength of the
Hund's coupling and the Coulomb repulsions, while the coexistence of
the field-induced SDW order and SC order around the core region is
realized due to the fact that the two orders emerge at different
energies. The LDOS at the core region for the vortex state with SDW
displays the dual structures with one reflecting the SC pairing and
the other being related to the SDW order. These features can be
discernable in the STM measurements for identifying the interplay
between the field-induced SDW order and the SC order around the core
region.

\section{acknowledgement}
\par This project was supported by
National Natural Science Foundation of China (Grant No. 10525415 and
No. 10904062), the Ministry of Science and Technology of Science
(Grants Nos. 2006CB601002, 2006CB921800), a GRF grant of Hong Kong
(HKU7055/09P), the China Postdoctoral Science Foundation (Grant No.
20080441039), and the Jiangsu Planned Projects for Postdoctoral
Research Funds (Grant No. 0801008C).

{\it Note added}-After completing this work, we get to know a
related work done by X. Hu, C. S. Ting, and J. X. Zhu~\cite{xhu}.

\vspace*{.2cm}
\begin{figure}[htb]
\begin{center}
\vspace{-.2cm}
\includegraphics[width=260pt,height=230pt]{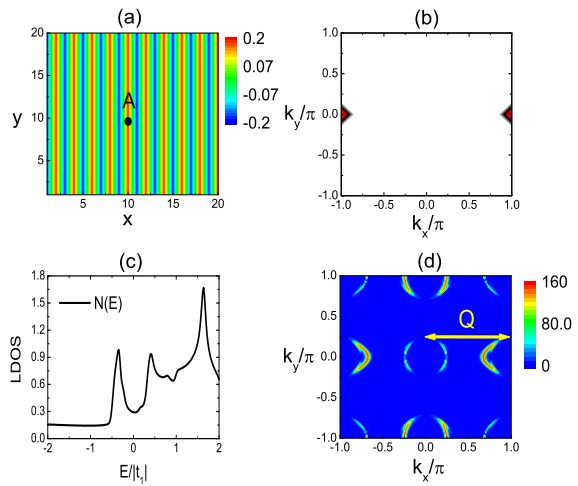}
\caption{(Color online) (a) The real space distribution of the
moment $M_{i}$. (b) The Fourier transformation of $M_{i}$. (c) The
LDOS at the site labeled as A in Fig. 1(a) in the SDW state. and (d)
spectral weight distribution in the SDW state (see
text).}\label{fig1}
\end{center}
\end{figure}
\vspace*{-.2cm}
\begin{figure}[htb]
\begin{center}
\vspace{-.2cm}
\includegraphics[width=270pt,height=130pt]{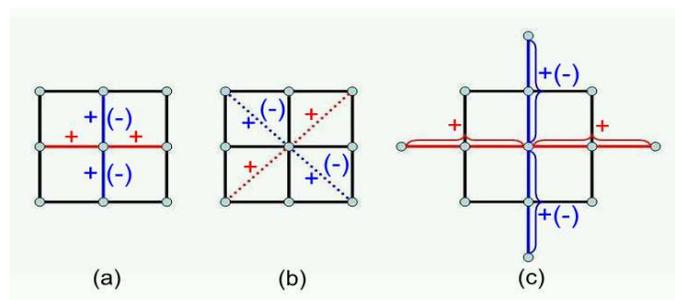}
\caption{(Color online) The SC order parameter configuration in the
real space between the NN sites pairing (a), the NNN sites pairing
(b), and the TNN sites pairing (c), respectively. The minus signs in
the parentheses correspond to the $d$-wave pairings, otherwise the
$s$-wave pairings.}\label{fig2}
\end{center}
\end{figure}
\vspace*{-.2cm}
\begin{figure}[htb]
\begin{center}
\vspace{-.2cm}
\includegraphics[width=260pt,height=230pt]{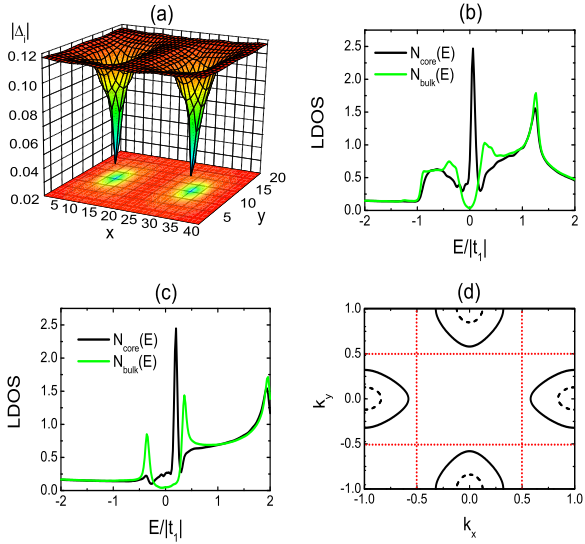}
\caption{(Color online) (a) The real space distribution of the SC
order amplitude $|\Delta_{i}|$ without the field-induced SDW. The
LDOS curves at the core center (black line) and for the bulk system
(green line) with electron doping $n=2.2$ (b), and hole doping
$n=1.8$ (c), respectively. (d) The electron-pocket in the
iron-pnictides in the unfolded Brillouin zone $-\pi\leq k_{x}<\pi$,
$-\pi\leq k_{y}<\pi$. The solid and dashed curves correspond to
electron doping with $n=2.2$ and hole doping with $n=1.8$,
respectively. The dotted (red) lines mark the nodal lines at
($¡À\pi/2, k_{y}$) and ($k_{x},¡À\pi/2$) for the
$s_{\pm}=\Delta_{0}\cos(k_{x})\cos(k_{y})$ order
parameter.}\label{fig3}
\end{center}
\end{figure}
\vspace*{-.2cm}
\begin{figure}[htb]
\begin{center}
\vspace{-.2cm}
\includegraphics[width=260pt,height=230pt]{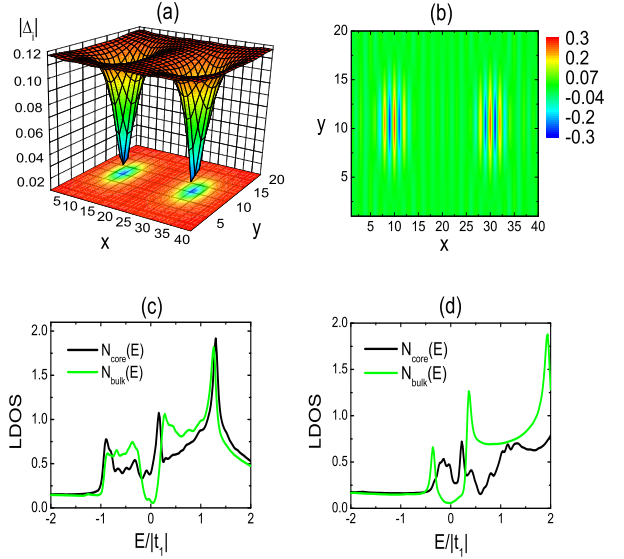}
\caption{(Color online) (a) The real space distribution of the SC
order amplitude $|\Delta_{i}|$ in the presence of the field-induced
SDW. (b) The real space distribution of the field-induced moment
$M_{i}$. The LDOS curves at the core center (black line) and for the
bulk system (green line) with electron doping $n=2.2$ (c), and hole
doping $n=1.8$ (d), respectively.}\label{fig4}
\end{center}
\end{figure}
\vspace*{-.2cm}
\begin{figure}[htb]
\begin{center}
\vspace{-.2cm}
\includegraphics[width=210pt,height=210pt]{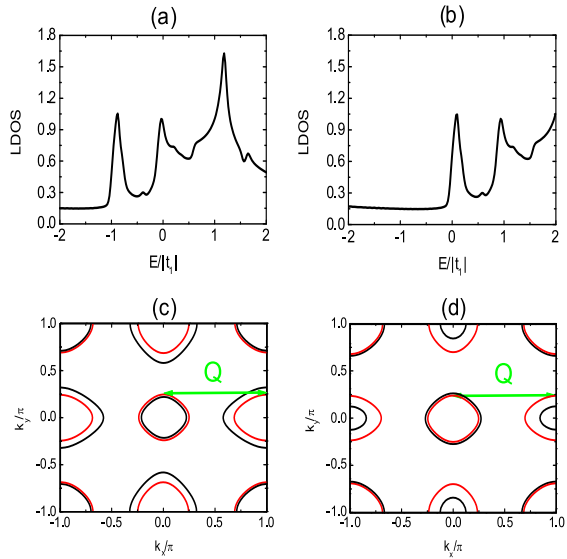}
\caption{(Color online) LDOS curves in a bulk system simulating the
SDW state around the core region for $n=2.2$ (a), and $n=1.8$ (b),
respectively. The wave vector $Q$ of the field-induced SDW connects
the finite energy contour (red line) for $n=2.2$ (c), and $n=1.8$,
respectively (Black curves denote the Fermi surface at corresponding
doping level).}\label{fig5}
\end{center}
\end{figure}
\vspace*{-.2cm}

\end{document}